\documentclass{iau}
\usepackage{graphicx}
\usepackage{natbib}
\usepackage{amsmath}

\usepackage[T1]{fontenc}
\usepackage{aecompl}

\usepackage[colorlinks, linkcolor={red}, citecolor={blue}, urlcolor={blue}]{hyperref}

\title[DIBs as probes of small-scale interstellar structure] 
{Diffuse interstellar bands as probes of small-scale interstellar structure}

\author[K. T. Smith, M. A. Cordiner \& P. J. Sarre]   
{Keith T. Smith$^{1,2}$\footnote{Email: {\tt kts@ras.org.uk}}, Martin A. Cordiner$^3$ \and Peter J. Sarre$^1$}

\affiliation{$^1$School of Chemistry, The University of Nottingham, Nottingham, NG7 2RD, UK\\[\affilskip]
$^2$Royal Astronomical Society, Burlington House, Piccadilly, London, W1J 0BQ, UK\\[\affilskip]
$^3$NASA Goddard Space Flight Center, 8800 Greenbelt Road, Greenbelt, MD 20770, USA}

\pubyear{2013}
\volume{297}  
\pagerange{xxx--yyy}
\date{?? and in revised form ??}
\jname{The Diffuse Interstellar Bands}
\editors{J. Cami \& N. L. J. Cox, eds.}
\begin{document}

\maketitle

\begin{abstract}
We present observations which probe the small-scale structure of the interstellar medium using diffuse interstellar bands (DIBs). Towards HD~168075/6 in the Eagle Nebula, significant differences in DIB absorption are found between the two lines of sight, which are separated by 0.25~pc, and $\lambda5797$ exhibits a velocity shift. Similar data are presented for four stars in the $\mu$~Sgr system. We also present a search for variations in DIB absorption towards $\kappa$~Vel, where the atomic lines are known to vary on scales of $\sim10$~AU. Observations separated by $\sim9$~yr yielded no evidence for changes in DIB absorption strength over this scale, but do reveal an unusual DIB spectrum.
\end{abstract}

\firstsection 
\section{Introduction}

Early theories of the diffuse interstellar medium predicted that there should be no significant structure on scales below $\sim 1$~pc. However, observations have shown that there can be large variations in interstellar column densities between lines of sight separated by as little as $\sim 10$~AU ($5 \times 10^{-5}$~pc).

Previous observations have relied on narrow atomic and molecular lines, requiring high-resolution observations at high signal-to-noise, which is very demanding in terms of telescope time. DIBs have recently been shown to exhibit similar variations on small scales \citep{Cordiner2006,Cordiner2013}, and can be observed at lower resolution. They can be used to identify promising lines of sight for high-resolution follow-up and to probe the spatial distribution of the DIB carriers. By comparing the variations seen in different DIBs, it may also be possible to further constrain theories of the carriers.

\section[mu Sagittarii]{$\boldsymbol\mu$~Sagittarii}

\begin{figure}[tb]
 \centering
 \includegraphics[width=0.70\textwidth]{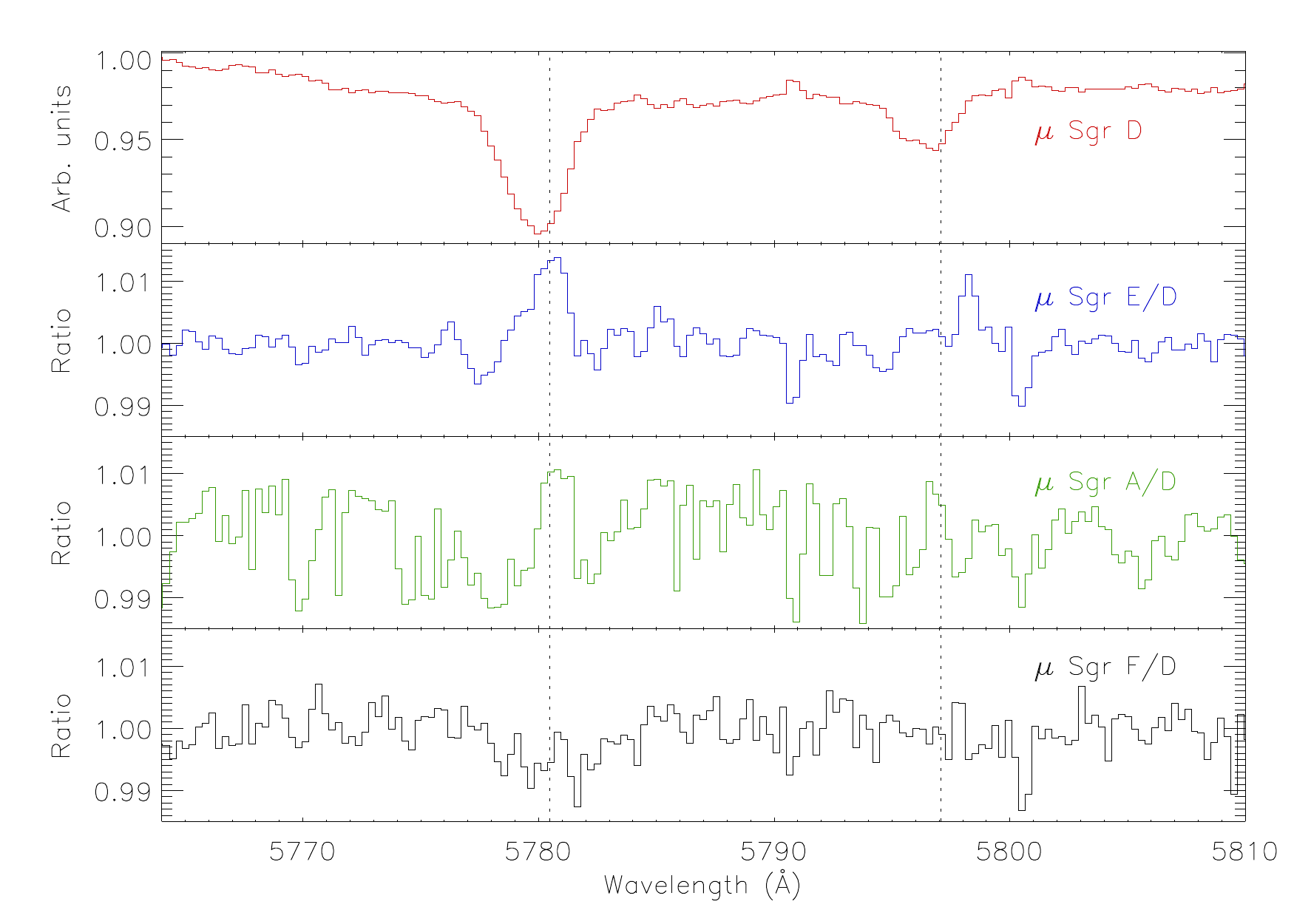}
 \caption{Comparison of the $\lambda5780$ and $\lambda5797$ DIBs towards four stars of the $\mu$~Sgr system. Top panel shows the DIBs towards $\mu$~Sgr~D, whilst the other three panels are the spectra of $\mu$~Sgr~E, A and F. respectively, each divided by the spectrum of $\mu$~Sgr~D. Dotted lines illustrate the DIB rest wavelengths from \citet{Hobbs2008}. Reproduced from \citet{Smith2010}, with permission.}
   \label{fig:mSgr}
\end{figure}

As a proof of concept, we observed DIBs towards four stars in the $\mu$~Sgr system ($\mu$~Sgr~A, D, E and F). Using the Robert Stobie Spectrograph (RSS) on the Southern African Large Telescope (SALT) at a resolving power $R\sim5,000$, we were able to obtain high quality spectra of DIBs even in short integration times. In Fig.~\ref{fig:mSgr} Significant differences are seen between DIBs towards $\mu$~Sgr~E and D, which are separated by $\sim0.5$~pc. Full results are presented in \citet{Smith2010}.

\section{HD~168075/6}

HD~168075 and HD~168076 are a pair of O-type stars which provide most of the UV photons that excite the Eagle Nebula. Using RSS on SALT, we observed HD~168075/6 at a resolving power of $R\sim8,000$, with both stars on the slit simultaneously. The integration time was just 570~s, despite thin cloud, poor seeing, and a narrow ($0.6''$) slit.

\begin{figure}[tb]
 \centering
 \includegraphics[width=0.70\textwidth]{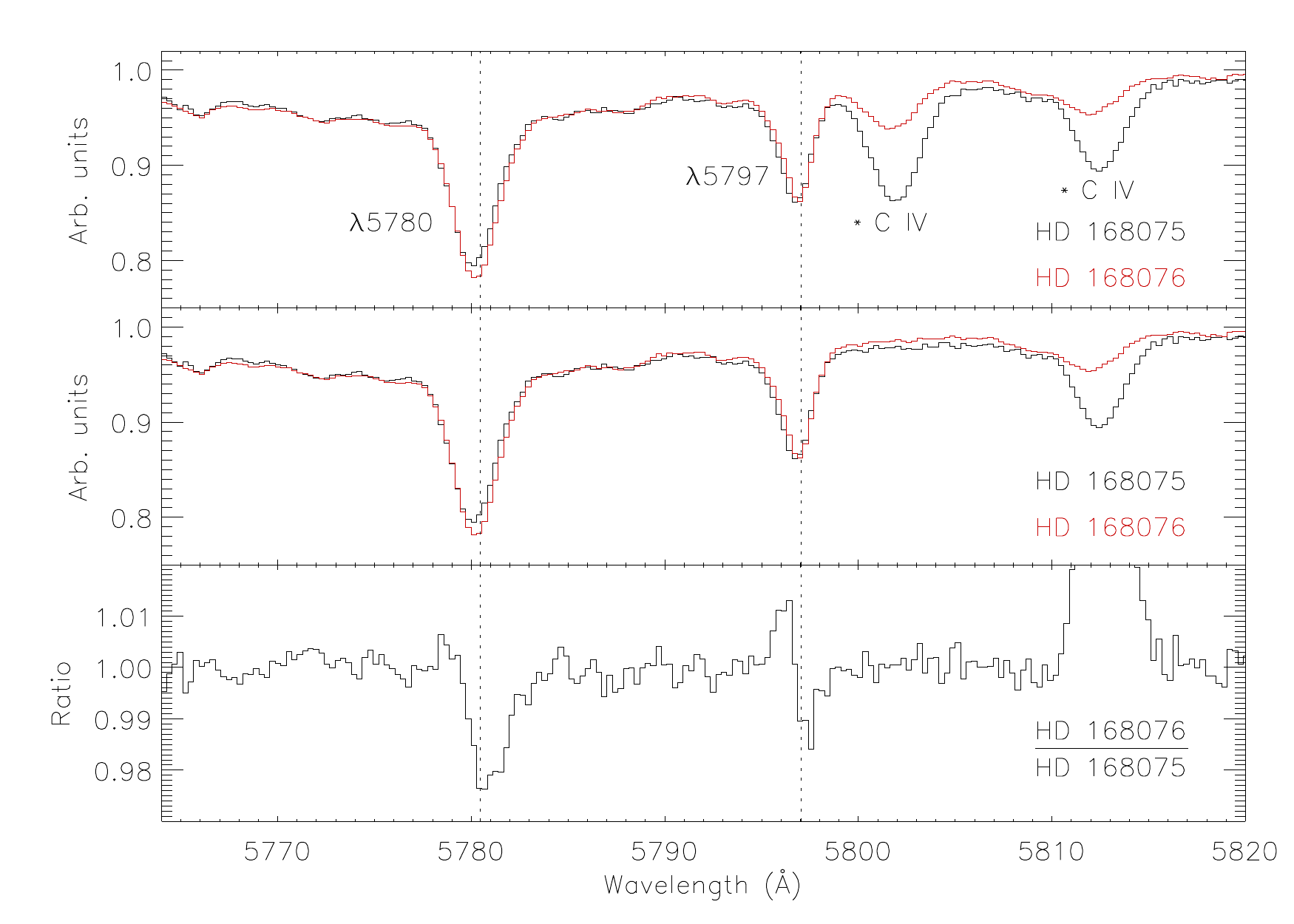}
 \caption{Comparison of spectra of HD~168075 and HD~168076 covering the $\lambda5780$ and $\lambda5797$ DIBs. Upper panel: reduced spectra. Middle panel: after removal of the stellar C~\textsc{iv} line at 5802~\AA\ to provide a clean continuum. Lower panel: ratio of the two spectra. Dotted lines illustrate the DIB rest wavelengths from \citet{Hobbs2008}. $\lambda5780$ is stronger towards HD~168076, whilst $\lambda5797$ exhibits a velocity shift. Reproduced from \citet{Smith2010}, with permission.}
   \label{fig:hd168}
\end{figure}

Fig.~\ref{fig:hd168} shows the $\lambda5780$ and $\lambda5797$ DIBs towards HD~168075/6. There are highly significant variations between the two lines of sight. $\lambda5780$ is stronger towards HD~168076, whilst $\lambda5797$ appears to shift in velocity. Full results for the $\lambda\lambda5780$, 5797, 5850, 6196, 6203, 6269, 6283 and 6613 DIBs are presented in \citet{Smith2010}.

\section[kappa Velorum]{$\boldsymbol\kappa$~Velorum}
\begin{figure}[tb]
 \includegraphics[width=0.55\textwidth]{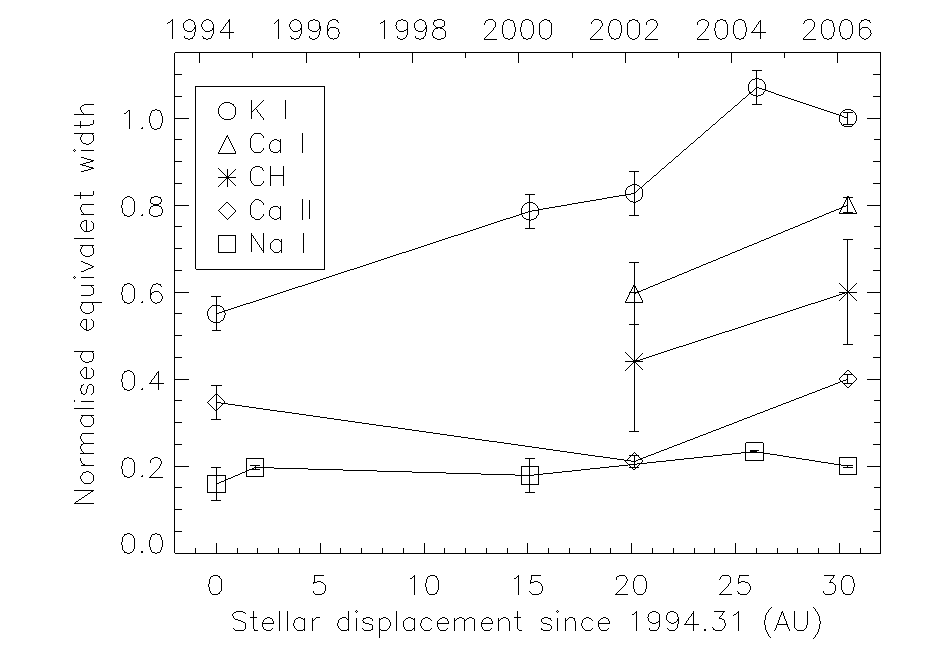} \includegraphics[width=0.45\textwidth]{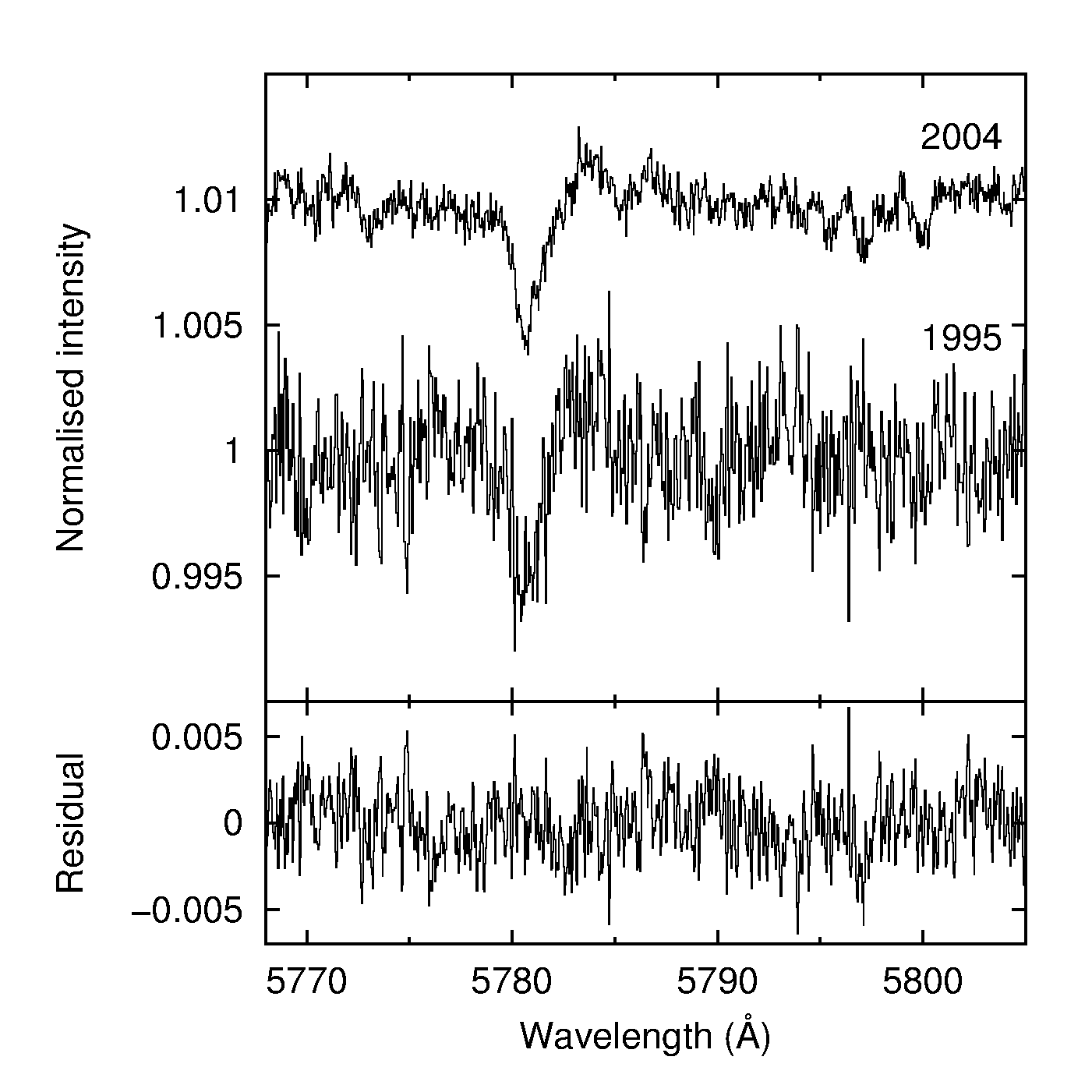}
 \caption{Left: Changes in the equivalent widths of atomic and molecular lines towards $\kappa$~Vel between 1994 and 2006. Right: The $\lambda5780$ DIB towards $\kappa$~Velorum in 1995 and 2004. $\lambda$5797 is not detected at either epoch, and there is no evidence for changes in $\lambda5780$ between the two epochs. Both panels reproduced from \citet{Smith2013}, with permission.}
   \label{fig:kVel}
\end{figure}

The interstellar absorption lines towards $\kappa$~Velorum are known to be time variable. For example, the equivalent width of the K~\textsc{i} 7698~\AA\ line has increased by $82\pm7\%$ between 1994 and 2006, during which proper motion of the star has carried it 30~AU perpendicular to the line of sight (Fig.~\ref{fig:kVel}).

Using observations from Mount Stromlo in 1995 and the Anglo Australian Telescope in 2004, we searched for variations in the DIBs towards this star. None were found, with an upper limit of $<40\%$ on changes in $\lambda5780$ (Fig.~\ref{fig:kVel}). The $\lambda5780/\lambda5797$ equivalent width ratio observed in 2004 is at least $> 22$ ($1\sigma$), which is the highest ever reported. Full results have been published in \citet{Smith2013}.

\section{Conclusions}
Significant differences in DIB absorption have been found towards the $\mu$~Sgr and HD~168075/6 systems, on scales of less than a parsec. These results confirm the discovery by \citet{Cordiner2006,Cordiner2013} that DIBs can be used as probes of small-scale interstellar structure, and demonstrate that moderate-resolution observations of DIBs may be used to identify promising targets for high-resolution studies. Although the observations of $\kappa$~Vel found no variation in DIBs on scales of $\sim30$~AU, the DIB spectrum itself was found to be unusual, with a very high $\lambda5780/\lambda5797$ ratio.


\begin{thebibliography}{}

\bibitem[Cordiner \etal(2006)]{Cordiner2006}
{Cordiner, M.~A., Fossey, S.~J., Smith, A.~M., Sarre, P.~J.} 2006, \textit{Faraday Discussions}, 133, 403

\bibitem[Cordiner \etal(2013)]{Cordiner2013}
{Cordiner, M.~A., Fossey, S.~J., Smith, A.~M., Sarre, P.~J.} 2013, \textit{ApJ}, 764, L10

\bibitem[Hobbs \etal(2008)]{Hobbs2008}
{Hobbs, L.~M. \etal} 2008, \textit{ApJ}, 680, 1256

\bibitem[Smith(2010)]{Smith2010}
{Smith, K. T.} 2010, PhD thesis, The University of Nottingham

\bibitem[Smith \etal(2013)]{Smith2013}
{Smith, K.~T., Fossey, S.~J., Cordiner, M.~A., Sarre, P.~J., Smith, A.~M., Bell, T.~A. \& Viti, S.} 2013, \textit{MNRAS}, 429, 939

\end{thebibliography}
\end{document}